\documentclass[pra,aps,onecolumn,superscriptaddress,nofootinbib]{revtex4}

\usepackage{hyperref}
\usepackage{color}
\usepackage{url}

\usepackage{amsmath}
\usepackage{amsfonts}
\usepackage{amssymb}
\usepackage{bm}
\usepackage{bbm}
\usepackage{nicefrac}
\usepackage{xcolor}
\usepackage{pifont}
\usepackage{graphicx}
\usepackage{enumerate}
\usepackage{dcolumn}
\usepackage{maybemath}

\def\lambdabar{{\mskip2mu\mathchar'26\mkern-9.8mu\lambda}}

\newcommand{\dd}{\mathrm d}
\newcommand{\ee}{\mathrm e}
\newcommand{\ii}{\mathrm i}

\newcommand{\calL}{\mathcal L}
\newcommand{\calO}{\mathcal O}

\definecolor{darkgreen}{rgb}{0,0.42,0.23}

\begin{document}


\title{Attempts at a determination of the fine-structure constant
from first principles:\\
A brief historical overview}

\author{U. D. Jentschura}

\affiliation{Department of Physics,
Missouri University of Science and Technology,
Rolla, Missouri 65409-0640, USA}

\affiliation{MTA--DE Particle Physics Research Group,
P.O.Box 51, H--4001 Debrecen, Hungary}

\author{I. N\'andori}

\affiliation{MTA--DE Particle Physics Research Group,
P.O.Box 51, H--4001 Debrecen, Hungary}


\begin{abstract}
It has been a notably elusive task to find a remotely sensical ansatz
for a calculation of Sommerfeld's electrodynamic fine-structure constant
$\alpha_{\rm QED} \approx 1/137.036$ based on first principles. However, this
has not prevented a number of researchers to invest considerable effort into
the problem, despite the formidable challenges, and a number of attempts have
been recorded in the literature.  Here, we review a possible approach based on
the quantum electrodynamic (QED) $\beta$ function, and on algebraic identities
relating $\alpha_{\rm QED}$ to invariant properties of ``internal'' symmetry
groups, as well as attempts to relate the strength of the electromagnetic
interaction to the natural cutoff scale for other gauge theories.  Conjectures
based on both classical as well as quantum-field theoretical considerations are
discussed.  We point out apparent strengths and weaknesses of the most
prominent attempts that were recorded in the literature.  This includes
possible connections to scaling properties of the Einstein--Maxwell Lagrangian
which describes gravitational and electromagnetic interactions on curved
space-times. Alternative approaches inspired by string theory are also
discussed.  A conceivable variation of the fine-structure constant with time
would suggest a connection of $\alpha_{\rm QED}$ to global structures of the
Universe, which in turn are largely determined by gravitational interactions.\\[2ex]
{{\bf 11.25.Tq}}{\,(Gauge field theories)\;;}\\
{{\bf 11.15.Bt}}{\,(General properties of perturbation theory)\;;}\\
{{\bf 04.60.Cf}}{\,(Gravitational aspects of string theory)\;;}\\
{{\bf 06.20.Jr}}{\,(Determination of fundamental constants)\,.}
\end{abstract}


\maketitle



%
%
\section{Introduction}
\label{sec1}

Today, the determination of a viable analytic formula for the fine-structure
constant remains an extremely elusive problem. The fine-structure constant
$\alpha_{\rm QED} \approx 1/137.036$  is a dimensionless physical constant, and
any conceivable variation of it with time~\cite{Di1938,FiEtAl2004} would
necessarily imply a connection of electromagnetic interactions to other global
properties of the Universe, such as its age. Alternatively, one may  point out
that expressions for the fine-structure constant in terms of well-defined
mathematical invariants of an underlying symmetry group~\cite{Wy1969,Wy1971}
are incompatible with a variation of the fine-structure constant with time
(unless the symmetry group changes with time also).  Indeed, the problem of
finding a conceivable analytic formula for the fine-structure constant is of
such fundamental interest that considerable field-theoretical insight and
effort has been invested into the task, despite the formidable challenges.  It
thus appears useful to review the historical development and status of these
attempts, and to indicate possible future directions of research, while noting 
that considerable scrutiny and scepticism are appropriate with regard to 
the elusive and formidable challenge.

Let us start by recalling that the quantum electrodynamic (QED) $\beta$
function~\cite{GMLo1954,BaJo1969,AkBe1969} describes the evolution of the QED
running coupling $\alpha_{\rm QED}$ over different momentum scales; one may
naturally ask the question if the physical value of $\alpha_{\rm QED}$ is
related to a specific momentum scale. Indeed, QED is first and foremost defined
at a high-energy scale (cutoff scale), and the renormalization-group (RG)
evolution of $\alpha_{\rm QED}$ can thus be used to evolve the coupling into
the low-energy domain. If one postulates certain constraining properties of
$\alpha_{\rm QED}$ either in the high-energy, or the low-energy, limit, then
one might hope~\cite{Ad1972,Ma1984rg} to obtain a constraint equation which
determines a physically reasonable approximation to $\alpha_{\rm QED}$.
However, the so-called triviality of QED~\cite{ZJ2007} poses a very interesting
question for physicists, namely, to explain the numerical value of $\alpha_{\rm
QED}$ without having, as an ``anchor point'', a ``critical value'' for the
coupling: From the point of view of renormalization-group (RG)
theory~\cite{ZJ2007}, QED does not have a phase transition. Consequences of
these observations are discussed in Sec.~\ref{sec2A}.

Another intuitive ansatz for the determination of $\alpha_{\rm QED}$ would a
priori involve algebraic considerations which relate $\alpha_{\rm QED}$ to
certain invariants of internal symmetry groups, e.g., those which describe the
intrinsic spin of a Dirac particles. Other attempts are based on possible
connections of the QED coupling to invariants of higher-dimensional internal
symmetry groups, whose projection onto four dimensions yields a value for
$\alpha_{\rm QED}$ close to the observed parameter.  Related attempts are
discussed in Sec.~\ref{sec2B}.

However, one may point out that ``stand-alone'' approaches to the calculation
of $\alpha_{\rm QED}$, discussed in Sec.~\ref{sec2}, would necessarily imply
that the value of $\alpha_{\rm QED}$ is determined by invariants of an
underlying symmetry group. (In the case of the QED $\beta$ function, the $U(1)$
gauge group determines the coefficients of the $\beta$ function.)  Under these
assumptions, the symmetry group fixes the value of the fine-structure constant,
and $\alpha_{\rm QED}$ will remain constant with time. 
(This of course does not hold 
if the symmetry group changes with time, i.e., if QED suddenly
evolves from a $U(1)$ gauge theory to a different internal symmetry group.)
Dirac's large number hypothesis~\cite{Di1938} conjectures 
that $\alpha_{\rm QED}$ changes with the age of the Universe, and 
a number of recent papers~\cite{WeFlChDrBa1999,DzFlWe1999,%
WeEtAl2001,MuWeFl2003,WeEtAl2011} claim to have established a variation
of $\alpha_{\rm QED}$ with time (see Sec.~\ref{sec3}).

The large-scale structure of the Universe is determined first and foremost by
gravitational interactions. This has inspired some researchers to look for
connections of quantum fluctuations of the quantum fields on the curved
space-time, and possible connections of the effective Lagrangian obtained after
the integration of the ``heavy'' degrees of freedom, to gravity (see
Sec.~\ref{sec4}).  Indeed, if there is a variation of $\alpha_{\rm QED}$
connected to changes in the global properties of the Universe (such as its age
which trivially evolves with time), then a change in the strength of the
electromagnetic interaction suggests a connection of the electrodynamic and
gravitational interactions, because the latter in turn determine the global
properties of the Universe (see Sec.~\ref{sec5}).  One of the first attempts at
a unification of two fundamental forces, namely, gravitation and
electromagnetism, involves a five-dimensional generalization of space-time
(Kaluza--Klein theories, see Refs.~\cite{Ka1921apsrmp,Kl1926,OvWe1998}). This ansatz
involves a compactification of the fifth dimension which in turn leads to a
natural charge quantization condition, and results in a formula connecting the
fine-structure constant with a background field (see Sec.~\ref{sec5B}).  It is
also instructive to recall (see Sec.~\ref{sec5A}) that the known, manifestly
nonvanishing photon-graviton conversion amplitudes establish a connection of
gravity and electromagnetism within a fully quantized formalism.

An alternative attempt at analyzing a connection of
gravity and electromagnetism is inspired by 
string theory~\cite{Po1998vol1,Po1998vol2}.
Roughly speaking, the scattering amplitudes derived from open as opposed to closed
strings strongly suggest that a connection of $\alpha_{\rm QED}$ and
$\alpha_G$ (the fine-structure constant of gravitational interactions) 
might exist. Furthermore, the functional relationship suggests that 
$\alpha_{\rm QED}$ (related to an open-string amplitude) should be
proportional to the square root $\sqrt{\alpha_G}$ of the gravitational
fine-structure constant (closed-string amplitude),
modulo a proportionality factor that might depend on other 
fundamental constants. Possible consequences of
this observation are discussed in Sec.~\ref{sec5C}. 

We frequently use the
Newtonian gravitational constant $G$, Planck's unit of action $\hbar$, the
vacuum permittivity $\epsilon_0$, the electron and proton masses $m_e$ and
$m_p$, and the speed of light $c$.  SI mksA units 
are employed throughout the paper.

%
%
\section{Algebraic and Analytic Approach to the Fine--Structure Constant}
\label{sec2}

%
%
\subsection{Renormalization Group and the Fine--Structure Constant}
\label{sec2A}

It is very interesting to explore possible explanations of 
the observed relation $\alpha_{\rm QED}
\approx 1/137.036$ to the RG evolution of the coupling constant of quantum
electrodynamics (QED) with the momentum scale.  We recall that the RG evolution
of $\alpha_{\rm QED}$ is described by the
equation~\cite{GMLo1954,AkBe1969,BaJo1969,dRRo1974,ItZu1980,GoKaLaSu1991}
\begin{equation} \label{RG} \int_{G^{\rm as}(\alpha, x_1)}^{G^{\rm as}(\alpha,
x_2)} \frac{\dd z}{\beta(z)} = \ln\left( \frac{x_2}{x_1} \right) \,,
\end{equation}
where $G^{\rm as}(\alpha, x)$ is the running QED coupling in the asymptotic,
high-energy region, and $\alpha$ is its reference value at the momentum scale
$x$ where the RG evolution starts (i.e., where the coupling is ``matched''
against the physical value of $\alpha$).  Of course, the assumption is that an
analytic continuation of the QED $\beta$ function to the non-asymptotic,
low-energy region exists which allows us to use the functional form~\eqref{RG}
in the low-energy region. Indeed, the RG equation of Callen and
Symanzik~\cite{Ca1970,Sy1970} clarifies that, in the low-energy region, the
``running'' of the mass parameters cannot be ignored, but we ignore this
possible complication in the following discussion.  Pertinent remarks on the
``running'' of the mass parameter can be found in Chap.~13 of
Ref.~\cite{ItZu1980}, Refs.~\cite{ZJ2002,ZJ2007}.  For a discussion of the mass
running within QED, we refer~\cite{DuGiSc2002}.

Let us now assume that we are evolving the QED coupling 
downward with respect to the momentum scale.
We assume that $x_1 \to \infty$ defines a large
momentum scale, e.g., the momentum scale of the bare theory,
while we also assume that $x_2$ defines the scale where
the coupling $G^{\rm as}(\alpha, x_2) = \alpha_{\rm QED}$.
For $x_1 \to \infty$, the 
QED coupling describes the ``bare'' charge.
If we plainly enter with $x_1 \to \infty$ into 
the right-hand side of Eq.~\eqref{RG},
then it diverges logarithmically.
So, if QED is assumed to be valid 
across all momentum scales, then the only 
way to make the left-hand side of Eq.~\eqref{RG} 
also diverge logarithmically is to assume that 
\begin{equation}
\beta(G^{\rm as}(\alpha, x_2)) 
= \beta(\alpha^*) 
\mathop{=}^{\mbox{?!}} \beta(\alpha_{\rm QED})
\mathop{=}^{\mbox{!}} 0 \,,
\end{equation}
because the integrand $1/\beta(z)$ in Eq.~\eqref{RG} 
would in this case also diverge near 
$z \approx \alpha^* = G^{\rm as}(\alpha, x_2) = \alpha_{\rm QED}$.
This consideration motivates the conjecture~\cite{Ad1972,Ma1984rg} that 
$\alpha_{\rm QED}$ constitutes a zero of the QED $\beta$ function,
where we recall that the asterisk in the 
superscript $\alpha^*$ is used to 
denote a generic critical point of the RG evolution~\cite{ZJ2007}.
It has been pointed out in Ref.~\cite{Ad1972fermilab}
that this argument also holds if the QED $\beta$ function 
is restricted to subclasses of 
diagrams with only one closed fermion loop.
Indeed, the replacement 
$\beta \to F^{[1]}$ advocated in Ref.~\cite{Ad1972fermilab}
singles out the one-fermion-loop diagrams
(where $F^{[1]}$ singles out the one-loop 
diagrams) but does not change the overall physical picture discussed below.

We should also be careful, because we have 
in fact used the RG evolution equation in a region which is manifestly 
non-asymptotic with respect to the momentum scale, namely,
in a region where the running of the coupling cannot 
be separated from the running of the mass~\cite{ItZu1980}.
A more serious objection comes from the fact that
the conjecture is in obvious disagreement
with the concrete numerical values of the first
two perturbative coefficients of the QED $\beta$ function
(in the momentum scheme, see Ref.~\cite{GoKaLaSu1991}),
\begin{equation}
\label{betaMOM}
\beta(\alpha) =
\frac{\alpha^2}{3 \pi}
+ \frac{\alpha^3}{4 \pi^2}
+ \frac{\alpha^4}{\pi^3} \,
\left( \frac13 \, \zeta(3) - \frac{101}{288} \right)
+ \frac{\alpha^5}{\pi^4} \,
\left( -\frac53 \, \zeta(5) + \frac13 \, \zeta(3) + \frac{93}{128} \right) + 
\calO(\alpha^6) \,.
\end{equation}
While the coefficients of order $\alpha^2$ and $\alpha^3$ are both positive and
thus exclude a nontrivial positive of $\beta(\alpha)$ for small and
positive $\alpha$, one could argue that this says nothing about the large-order
(or strong-coupling) asymptotics of the $\beta$ function.  The apparent absence
of a zero of the QED $\beta$ function,
based on a consideration of the first perturbative 
coefficient, is also known as the ``triviality'' of
QED (see Ref.~\cite{LaAbKh1954}). 
Recently, using lattice calculations and exact RG approaches, the
triviality of QED (from the RG point of view, i.e., the 
absence of a phase transition) 
has been confirmed in Refs.~\cite{KiKoLo2002,GoEtAl1997,GiJa2004}.

One might still speculate about possible zeroes of the QED $\beta$ function in
the high-momentum region.  However, analytic arguments~\cite{Su2001,Su2009}
based on a rather sophisticated extrapolation of the perturbative approach of
the QED $\beta$ function to the nonperturbative (strong-coupling) domain
suggest that the value of the QED $\beta$ function increases~\cite{Su2009} with
the momentum scale, with $\beta(g) \propto g$ in the asymptotic region.  While
the linear increase with the momentum scale, if confirmed, would constitute a
rather surprising functional dependence, one may otherwise remark that the
result of Ref.~\cite{Su2009} is otherwise consistent with the absence of zeros
of the QED beta function. The linear increase  $\beta(g) \propto g$ of the QED
coupling with the momentum scale, proposed in Ref.~\cite{Su2009}, appears to be
at variance with the logarithmic dependence suggested by the dominant one-loop
evolution~\cite{ItZu1980}, and also, at variance with the results of a
sophisticated effective-charge approach~\cite{BiBr2003}.  However, the precise
functional form of the monotonous increase of the QED coupling does not really
matter: The QED $\beta$ function, as it follows from the RG program applied at
face value to QED, is universally assume to remain positive over all momentum
scales, thus excluding a nontrivial zero.

One may also explore the possibility of an ultraviolet (UV)
fixed point of QED, 
with $\beta(\alpha = \alpha^*) = 0$ in the UV,
where the QED $\beta$ function in the on-mass-shell scheme 
runs into this fixed point in the ultraviolet, i.e.,
\begin{equation}
\alpha_\infty = G^{\rm as}(\alpha^*, x \to \infty) \,,
\qquad
\beta(\alpha^*) = 0 \,.
\end{equation}
In this case, the effective QED coupling should attain a well-defined, finite
value at infinite momentum transfer, and $\alpha^*$ would have to be
interpreted as the finite bare charge.
This assumption is at variance with predictions of the 
Landau pole theory~\cite{LaAbKh1954,ItZu1980,Su2001,Su2009}, 
which would predict the QED coupling to diverge at an intermediate momentum scale
$q^2_*$ [interestingly, this pole persists even at
finite temperature, see Refs.~\cite{JaMa2012,JaMa2013,JaMa2014}],
and also, with more general arguments~\cite{Su2001,Su2009}
that suggest a monotonous increase of the QED coupling at high momenta (without
intermediate poles of the Landau type).

Let us conclude this discussion by observing that the 
QED $\beta$ function has an (almost trivial) zero at
the Gaussian fixed point,
\begin{equation}
\beta( \alpha^* ) = \beta( \alpha^* = 0 ) = 0 \,.
\end{equation}
(The notation $\alpha^*$ is normally reserved for the first nontrivial zero of
the QED $\beta$ function, i.e., for a nonvanishing value of $\alpha^*$, but we
extend the notation here to the value $\alpha^* = 0 $, which is the trivial
Gaussian fixed point.) The famous argument of Dyson~\cite{Dy1952} identifies
the point $\alpha^* = 0$ (transition from positive to negative QED coupling) as
the point where the vacuum becomes unstable against the creation of
electron-positron pairs, due to quantum tunneling. 
In that sense, the value $\alpha^* = 0$ constitutes a
critical point of the RG evolution of the QED coupling,
but it does not match the physical value of $\alpha_{\rm QED}$
which is manifestly nonvanishing.

The dependence on the order of perturbation theory of the first coefficients of
the QED $\beta$ function in Eq.~\eqref{betaMOM} does not exhibit the asymptotic
(factorially divergent) structure which is otherwise useful in determining
critical exponents as a function of the critical coupling, e.g., in the
$N$-vector model~\cite{GuZJ1998}.  We have carried out, independently,
numerical attempts to determine an estimate for the first positive or negative
zero $\alpha^*$ of the QED $\beta$ function, based on the coefficients given in
Eq.~\eqref{betaMOM}. These attempts have
been unsuccessful (here, the ``first zero'' is to be
interpreted as the one with minimum modulus, for either positive 
or negative $\alpha_{\rm QED}$). This conclusion remains valid even if 
Pad\'{e} approximants are used
in order to improve the convergence of the perturbative expansion of the
$\beta$ function. Furthermore, the zero of the 
RG evolution, determined by the first two coefficients
of Eq.~\eqref{betaMOM},
\begin{equation}
\beta(\alpha^{*}) \approx
\frac{(\alpha^*)^2}{3 \pi}
+ \frac{(\alpha^*)^3}{4 \pi^2} \mathop{=}^{\mbox{!}} 0 \,,
\qquad
\alpha^* = -\frac{4 \pi}{3} \approx -4.189 \,,
\end{equation}
lies in the ``unphysical'', ``unstable''
region where the QED vacuum becomes unstable against the 
creation of electron-positron pairs (which would repel 
each other, so that the vacuum energy can be lowered
by separating the charges into distant regions
of space, still conserving electric neutrality of the 
Universe, see Ref.~\cite{Dy1952}). Also, since the modulus 
of $\alpha^* = -4.189$ is of order unity (or, larger than unity),
one may expect that the first perturbative terms from Eq.~\eqref{betaMOM}
cannot give a reliable estimate for its numerical value.
However, one may speculate about the physical significance of 
$\alpha^* = -4.189$ (or, of an improved determination
thereof based on higher-order perturbative terms)
as follows: Namely, for negative $\alpha$ of unit
modulus, the binding energy of two (mutually attracting, in this case) 
electrons, or two (mutually attracting) positrons becomes 
commensurate with their rest energy.
The negative critical value $\alpha^* = -4.189$ might thus describe 
a phase transition where the spontaneous pair
creation from the vacuum, which sets in immediately 
at $\alpha = 0$ (branch point of the QED perturbation series),
in fact turns into spontaneous creation of electron-positron
pairs in bound as opposed to free states (i.e., pair 
creation into bound ``Cooper pairs''
consisting of either two mutually attracting electrons, or two 
mutually attracting positrons). In these modified ``Cooper pairs'',
the dominant attractive binding force would come from the strong, mutually attracting,
electrostatic Coulomb interaction (which is reversed in 
sign because we consider a hypothetical situation with $\alpha < 0$). 

All considerations reported in the current 
section are consistent 
with the absence of a phase transition of QED for any positive
value of $\alpha$, and with the lack of an explanation for the 
physically observed value of $\alpha_{\rm QED}$, based on the 
RG evolution of the QED coupling parameter.

%
%
\subsection{Algebraic Relations and the Fine--Structure Constant}
\label{sec2B}

The scientific literature is not free from attempts to determine $\alpha_{\rm
QED}$ based on algebraically simple combinations of transcendental numbers like
$\pi$, or logarithms of characteristic dimensionless physical quantities, which
approximate the numerical value of the QED coupling~$\alpha_{\rm QED} \approx
1/137.036$.  Other attempts are inspired by characteristic ratios in classical
``spin'' orbits of the electron~\cite{ScWi2012}.  The latter approach is
perhaps not totally unreasonable because the QED fine-structure constant
describes perturbation series for phenomena which are related to circular
motion.  E.g., in the semi-classical approximation, one may consider the Larmor
precession frequency of an electron in a magnetic field, which is proportional
to the electron (anomalous) magnetic moment, and yet, in its classical
analogue, is equivalent to the nutation frequency of the spinning electron
``top'' in the magnetic field. Still, the {\em classical} considerations
reported in Ref.~\cite{ScWi2012} have not been fully accepted as viable
explanations for an inherently {\em quantum} phenomenon, namely, the quantized
electromagnetic interaction. 

One may also counter-argue that it would be rather unusual if the precise
numerical value of a constant which is so intimately related to quantum
physics, such as $\alpha_{\rm QED}$, could be explained by invariant
characteristic numbers occurring in the classical analogue of the quantum
phenomenon.  The emergence of a dimensionless fundamental constant such
as $\alpha_{\rm QED}$ would naturally be assumed to be an inherent property of
the field theory which describes the underlying phenomena (namely, QED), rather
than a characteristic invariant of some quantum ``motion'' which is always
smeared out because of quantum fluctuations.  The approach discussed in
Refs.~\cite{ScWi2012} therefore appears doubtful.  Other numerical
coincidences fulfilled by the
fine-structure constant have recently been compiled in Ref.~\cite{Da2011}.

In principle, it would appear to be more promising to 
determine the fine-structure constant based on 
invariants of symmetry groups.  Wyler~\cite{Wy1969,Wy1971}
considers an a priori massless particle propagating in five-dimensional 
space, where the mass, in the fifth dimension,
according to the replacement $ m \to \ii \, \partial/\partial x_5$, 
parameterizes an internal time (or ``aging speed'') of the particle.
The invariance group $O(5,2)$ of the five-dimensional
Klein--Gordon equation is investigated in Refs.~\cite{Wy1969,Wy1971}.
The space ${\rm SO}(n,2)/[{\rm SO}(n) \otimes {\rm SO}(2)]$
relates the complex hypersphere $D^n$ and its characteristic boundary $Q^n$
to the spherical surface $S^{n-1}$.
Putting $n=5$, one obtains~\cite{HuMAGICapsrmp}
\begin{equation}
V(D^n) = \frac{\pi^n}{2^{n-1} \, n!} = \frac{\pi^5}{2^4 \, 5!} \,,
\quad
V(Q^n) = \frac{2 \pi^{n/2+1}}{\Gamma(n/2)} = \frac{2^3 \, \pi^3}{3} \,,
\quad
V(S^{n-1}) = \frac{2 \pi^{n/2}}{\Gamma(n/2)} = \frac{2^3 \, \pi^2}{3} \,.
\end{equation}
From these quantities, Wyler~\cite{HuMAGICapsrmp,Wy1969,Wy1971} assembles the following 
approximation to the fine-structure constant,
\begin{align}
\label{A1}
\alpha_{\rm QED} \approx & \;
2 \times 4\pi \, \frac{1}{V(S^4)} \, 
\frac{1}{V(Q^5)} \, \left[ V(D^5) \right]^{1/4} 
= \frac{9}{8 \pi^4} \, 
\left( \frac{\pi^5}{2^4 \, 5!} \right)^{1/4}
\nonumber\\[0.133ex]
=& \;  0.0072973\dots = \frac{1}{137.036\,0824\dots} \,.
\end{align}
One may argue that Wyler's considerations follow from the 
internal symmetry group of an equation describing 
a {\em noninteracting} particle (the {\em free}
Klein--Gordon equation, generalized to five dimensions), 
while QED necessitates the 
covariant coupling of the particle to the electromagnetic field operator
(and thus involves an {\em interaction} Hamiltonian).
Mathematically, Pease~\cite{Pe1977} argues that Wyler's 
calculation is based on an incorrect value of the 
coefficient of the Poisson kernel of certain five-dimensional domains 
considered in his formalism, and so the value of $\alpha_{\rm QED}$
may not be derivable from his assumptions.
Robertson~\cite{Ro1971} argues that the introduction 
of a additional scaling factor in Wyler's formula
destroys the (approximate) agreement with experiment.
Additionally, the numerical value derived from $\alpha_{\rm QED} $ 
in Eq.~\eqref{A1} is in disagreement with the latest
CODATA adjustment of the fundamental constants~\cite{MoTaNe2012}.

A different approach is taken by Rosen~\cite{Ro1976}
who assumes that the electromagnetic field operators
might be sums over $N=42$ ``hidden'' field operators,
where $N=42$ is the order of the transitive subgroup $G$ of the 
of the symmetric group of degree $7$, namely, $S_7$.
The physical value of $\alpha$ is determined 
by matching the vacuum expectation value of the 
sum over the ``hidden'' field 
operators, invariant under the internal symmetry 
group, against the ``effective'', ``physical'' QED field operator. 
Rosen~\cite{Ro1976} obtains the approximation
\begin{equation}
\label{A2}
\alpha_{\rm QED} \approx \frac{4 \pi}{N \, (N-1)} \approx 
\frac{1}{137.032\,406\dots} \,,
\qquad
N = 42 \,.
\end{equation}
However, both the choice of the order of the 
symmetry group ($S_6$ as well as $S_8$ are {\rm a priori} not excluded 
by physical considerations but yield markedly different 
value for $\alpha_{\rm QED}$).
Another main criticism is that the group-theoretical 
considerations leave almost no room for an ``adjustment'' of the 
fine-structure constant with regard to 
the observed deviations of the 
physical value of $\alpha_{\rm QED}$ from the 
values given in Eqs.~\eqref{A1} and~\eqref{A2}.
Either the values are exact, or group theoretical explanations
fail. At present, one should reemphasize that 
the value predicted for $\alpha_{\rm QED}$
by Eq.~\eqref{A2}, just like the value given in 
Eq.~\eqref{A1}, is incompatible with the CODATA value of the fine-structure 
constant~\cite{MoTaNe2012}.

%
%
\section{Conjectures Based on Classical Physics}
\label{sec3}

%
%
\subsection{Weyl's Hypothesis}
\label{sec3A}

In the year 1919, Weyl~\cite{We1919} formulated a conjecture 
relating the radius of the Universe $R_U$ and the classical
electron radius $r_e$,
\begin{equation}
r_e = \frac{\alpha \hbar}{m_e c} \,.
\end{equation}
He speculated that $r_e$ and $R_U$ 
might be related to a hypothetical ```radius'' of a particle whose rest mass
$m_H$ is equal to the naive expression for the
gravitational ``self-energy'' of the electron
(with dimension equal to its classical electron radius),
\begin{equation}
r_H = \frac{\alpha \hbar}{m_H c} \,,
\qquad
m_H \, c^2 = \frac{G m_e^2}{r_e} \,.
\end{equation}
The ratio of the two large quantities is observed to
be of the same order-of-magnitude,
namely 
\begin{equation}
\frac{r_H}{r_e} \sim 10^{42} \sim \frac{R_U}{r_e} \,.
\end{equation}
However, this observation is scrutinized by the
lack of a consistent interpretation of the classical
electron ``radius'' from a modern point of view.

%
%
\subsection{Dirac Large Number Hypothesis}
\label{sec3B}

Dirac's famous Large Number Hypothesis
dates back to the year 1938 (see Ref.~\cite{Di1938LNH}).
It is based on the observation that
the age $T$ of the Universe and the time it takes light to travel
a distance equal to the classical electron radius,
which is equal to $r_e/c$, are approximately proportional to each other,
\begin{equation}
\label{DiracT}
\frac{T}{(r_e/c)} \sim 10^{40} \sim
\frac{e^2}{4 \pi \epsilon_0 G \, m_e \, m_p}  \,.
\end{equation}
Dirac conjectured that the equality holds exactly,
and that the gravitational interaction constant $G$ might be 
inversely proportional to the age of the Universe $T$.

Alternatively, one might identify the expression $e^2/(4 \pi \epsilon_0)$ on
the right-hand side of Eq.~\eqref{DiracT} as $\alpha_{\rm QED} \, \hbar \,c$,
and conjecture that the fine-structure constant might be varying with the age
of the Universe. Indeed, such variations have been investigated recently by
Flambaum and others~\cite{WeFlChDrBa1999,DzFlWe1999,%
WeEtAl2001,MuWeFl2003,WeEtAl2011}. 
While observational data may indicate slightly lower values of $\alpha$
in the distant past (see Fig.~1 of Ref.~\cite{WeEtAl2001}),
a direct proportionality of the fine-structure constant 
to the age of the Universe has not been established conclusively.
In particular, in its most basic form, Dirac's hypothesis is incompatible 
with recent claims~\cite{WeEtAl2011} regarding a spatial 
variation (in addition to the temporal variation) 
of the fine-structure constant.

%
%
\subsection{Eddington Conjecture}
\label{sec3C}

Eddington~\cite{Ed1931} was the first to suggest a connection of the
gravitational and electromagnetic interactions, probably inspired by his
seminal work~\cite{Ed1924} on the theory of general relativity where, as well
shall see in the following, a certain connection of gravitational and
electromagnetic interactions is suggested by the structure of the Lagrangian.
Furthermore, as shown in Appendix~\ref{appa}, there exist certain analogies of
the gravitational and electromagnetic bound-state problems. The gravitational
fine-structure constant, defined in complete analogy with the electromagnetic
bound-state problem, depends on the masses of the two involved particles which
form the bound state and is typically much smaller than $\alpha_{\rm QED}$ for
the known elementary particles. Eddington conjectured (see Ref.~\cite{Ed1931})
that the electromagnetic fine-structure constant $\alpha_{\rm QED} \approx
1/137.036$ and the gravitational fine-structure constant $\alpha_G^{(ee)}$ for
two gravitationally interacting electrons should be proportional to each other,
\begin{align}
\alpha_{\rm QED} =& \; \frac{e^2}{4 \pi \epsilon_0 \hbar c} \,,
\qquad
\qquad
\alpha_G^{(ee)} = \frac{G \, m_e^2}{\hbar c} \,,
\\[0.133ex]
\label{edconj}
\frac{\alpha_{\rm QED}}{\alpha_G^{(ee)}} =& \;
\frac{e^2}{4 \pi \epsilon_0 G \, m^2_e} \approx 4.2 \times 10^{42}  
\approx \sqrt{ \, N_C \,} \,,
\end{align}
where $N_C$ is the number of charged particles in the Universe.  Eddington then
went on to conjecture that $N_C$ should be given explicitly in terms of an
integer number with $42$ decimals (the ``Eddington number''), invariant over
time, giving the number of positrons and electrons in the globally neutral
Universe. Invariably, the Eddington conjecture was formulated before the advent
of modern particle accelerators and lasers where particle creation processes
may be studied; e.g., a now rather famous experiment at the Stanford Linear
Accelerator Center (SLAC) described in Refs.~\cite{BuEtAl1997,BaEtAl1999} has
shown that electrons may be created in strong laser fields, after the injection
of a highly energetic $\gamma$ ray emitted by Compton backscattering from an
energetic oncoming electron.

%
%
\section{Quantum and Classical Fluctuations and the Fine--Structure Constant}
\label{sec4}

%
%
\subsection{Casimir Effect and the ``Mousetrap'' Model}
\label{sec4A}

The fine-structure constant is proportional to the ratio $e^2/(\hbar c)$. This
particular combination of physical quantities gives rise to a conjecture
formulated by Casimir~\cite{Ca1953}, as follows.  The charge distribution of a
truly elementary charged particle (like the electron, not the proton), in its
rest frame, is approximated as a conducting spherical shell carrying a
homogeneous surface charge of total magnitude $e$.  The electrostatic
self-energy of such an object is given as
\begin{equation}
E_S = \frac{e^2}{8 \pi \epsilon_0 \, a} \,,
\end{equation}
where $a$ is the radius of the shell.  This energy needs to be compared to the
Casimir energy of the spherical shell configuration, because vacuum
fluctuations of the electromagnetic field are influenced by the boundary
conditions set forth by the spherical shell configuration. 

E.g., according to the formalism of QED (see Chap.~3.2.4
of~Ref.~\cite{ItZu1980}), parallel plates are known to attract each other due
to the vacuum fluctuations; the latter are suppressed in the region in between the
plates, while they are not suppressed outside the plates, resulting in a net
attractive force.  Analogously, vacuum fluctuations are suppressed inside the
spherical shell.  Intuitively, one would assume that the vacuum fluctuations
outside the spherical shell, which are only marginally influenced by the
boundary conditions set forth by the (small) spherical shell of radius $a$,
would tend to press the spherical shell together, ``trapping'' the charge.
Hence, the name ``mousetrap model'' has been used in the
literature~\cite{Bo1968}.

Dimensional analysis shows that the 
Casimir energy of the spherical shell contribution 
has to be of the functional form
\begin{equation}
E_C = - C \, \frac{\hbar \, c}{2 a} \,,
\end{equation}
where $C$ is a constant to be determined.
Equating the surface tensions (or the restoring forces)
derived from the self-energy 
$E_S$ and the Casimir energy $E_C$, one 
arrives at the equilibrium condition
\begin{equation}
C = \frac{e^2}{\epsilon_0 \, \hbar \, c} \,.
\end{equation}
If we now assume that the model holds universally for all charged elementary
particles, then $C$ is promoted to the role of a dimensionless fundamental
constant of nature, exclusively determined by the Casimir configurations
(boundary conditions), and
proportional to the same combination of quantities $e$, $\hbar$, $\epsilon_0$
and $c$ as the fine-structure constant.  Furthermore, the formalism remains
valid in the limit $a \to 0$, thus avoiding discussion regarding the role of
$a$ as a yet-to-be-specified parameter of the ``mousetrap'' model (which 
thus would work for arbitrarily small ``mice'').

Unfortunately, closer inspection~\cite{Bo1968}
reveals that the most immediate ansatz for the shape of the 
fundamental charge distribution, namely, 
the spherical shell just discussed, leads to a 
repulsive rather than attractive Casimir energy~\cite{Bo1968},
thus invalidating (at least the most immediate version of)
the model.
However, one may point out that it would be very interesting 
to consider alternative shapes, and the last word on ``mousetrap 
models'' is yet to be spoken. Another aspect 
which would need to be taken into consideration is as follows.
While $C$ is related to the fine-structure 
constant, it is not necessarily equal to it, i.e.,
the model is derived from the self-interaction of a single
elementary particle, not from the interaction of 
different elementary particles, which characterizes the 
strength of the interaction.
Still, it might be worthwhile to explore variants of the 
``mousetrap model'' based on alternative shapes of the 
charge distribution in the future.

%
%
\subsection{Electromagnetic Fluctuations and Gravitational Interactions}
\label{sec4B}

Electrostatic interactions can either be attractive or repulsive, while gravity
always is attractive. However,  fluctuations of the electromagnetic field in
neutral objects typically lead to attractive interactions.  Examples are the
charge fluctuations of induced dipoles that lead to the van-der-Waals and
Casimir interactions among atoms~\cite{CaPo1948}.  Hence, it is tempting to
explore a possible ansatz that would identify gravitational interactions as
possible residual manifestations generated by  fluctuations electromagnetic (or
even electroweak) interactions.  While these approaches typically do not lead to
any concrete formula for the fine-structure constant, we still include a brief
discussion of related works in part because they naturally lead to the
field-theoretical conjectures discussed in Sec.~\ref{sec5}.

In Refs.~\cite{As1992,As1995}, one such possibility is explored.
It is shown that if every charged (or neutral) object is 
endowed with a fundamental set of spherically symmetric 
intrinsic electromagnetic oscillations, then these oscillations 
of the charge distributions will induce a force law 
at large distances which resembles Newton's gravitational 
force law, and which is always attractive. 
Gravity and electromagnetic interactions can then be unified 
provided a relation of the following type [see Eq.~(38) of 
Ref.~\cite{As1992}] holds universally for macroscopic bodies of 
masses $M_1$ and $M_2$,
\begin{equation}
\label{GMM}
G \, M_1 \, M_2 \propto 
\frac{q_{1+} \, A_{1-}^2 \, \omega_1^2 \,
q_{2+} \, A_{2-}^2 \, \omega_2^2}{4 \pi \epsilon_0} \,,
\end{equation}
the proportionality factor being of order unity.
Here, $q_{1+}$ and $q_{2+}$ are characteristic charges
of the microscopic entities in the two macroscopic bodies,
while $A_{1-}$ and $A_{2-}$ are conjectured to be 
of the order of an Angstrom ($10^{-10} \, {\rm m}$),
while the angular frequencies $\omega_1$ and $\omega_2$
are conjectured to be of order $10^9\,{\rm Hz}$. 
One may object, though, that the precise details of the 
material properties should otherwise enter the formalism,
and that it would be surprising if a general relationship
of the type given in Eq.~\eqref{GMM} could 
be established universally for all macroscopically
relevant physical samples,
relating the microscopic and macroscopic properties.
Nevertheless, the interpretation of gravitational interactions
are residual interactions stemming from fluctuations 
of fundamentally different physical origin is intriguing
and in fact, has been explored further.

One of the most prominent approaches in this 
direction has been given by Sakharov~\cite{Sa2000grg},
with a modern interpretation being supplemented by Visser~\cite{Vi2002}.
The basic idea is to study the quantum (as opposed to 
classical) fluctuations of fundamental quantum fields
(of the ``quantum vacuum''), and to try to relate these, 
on a macroscopic scale, to the gravitational interactions,
i.e., to space-time curvature.
The idea can roughly be formulated as follows:
One first assumes that the quantum fields 
should exist on a (possibly curved) Lorentzian manifold.
One does not attempt to quantize gravity itself,
but rather, one interprets gravity as being 
created as a residual interaction, due to the 
quantum fluctuations of the fields which ``live'' 
on the Lorentzian manifold.
Given a (possibly nonminimal) coupling of the 
quantum fields to the space-time curvature, 
it can be shown that the one-loop quantum fluctuations,
evaluated according to the Schwinger proper-time method~\cite{Vi2002},
give rise to terms of the form
\begin{equation}
\label{S1}
S = \int \dd^4 x \, \sqrt{- {\rm det} \, g} \,
\left( c_0 + c_1 \, R(g) + \dots\right) \,,
\end{equation}
where $c_0$ and $c_1$ are coefficient which are determined 
by the model, while the space-time curvature is $R(g)$, and 
the ellipsis ``$(\dots)$'' denotes higher-order correction
terms proportional to $R^2$. This action needs to be matched against 
the Einstein--Hilbert action (with a cosmological constant term),
\begin{equation}
\label{S2}
S = \int \dd^4 x \, \sqrt{- {\rm det} \, g} \,
\left( \Lambda + \frac{R(g)}{16 \pi \, G} + \dots\right)
\end{equation}
where again we ignore higher-order corrections
proportional to $R^2$. 

One possibility to generate the terms in Eq.~\eqref{S1}
is to consider the quantum fluctuations described by the 
one-loop effective action for a non-minimally coupled 
scalar field [see Eq.~(3) of  Ref.~\cite{Vi2002}],
\begin{equation}
S_g = -\frac12 \, \ln \, \det \left( \Box_g + m^2 + \xi \, R(g) \right)
\end{equation}
where $\Box_g$ is the ``quabla'' operator on the curved 
space-time manifold, while $m$ is the mass of the 
scalar field and $\xi$ is an effective coupling constant.
With a convenient reference metric $g_0$, one can write 
in the Schwinger proper-time representation
[see Chap.~4 of~\cite{ItZu1980} and Eq.~(3) of~\cite{Vi2002}],
\begin{equation}
S_g = S_{g_0} + \frac12 \, 
{\rm Tr} \,
\int_{\kappa^{-2}}^\infty \frac{\dd s}{s} 
\left[ \exp(-s [ \Box_g + m^2 + \xi \, R(g) ) 
- \exp(-s [ \Box_{g_0} + m^2 + \xi \, R(g_0) ) \right] \,,
\end{equation}
where $\kappa$ is a cutoff parameter in the 
proper-time. According to Eq.~(21) of Ref.~\cite{Vi2002},
and in agreement with arguments originally put forward by Sakharov~\cite{Sa2000grg},
the matching of $1/G$ in Eq.~\eqref{S2} against the quantum 
fluctuations leads to an expression which is 
quadratically divergent in $\kappa$. Thus, without recourse 
to further cancellations, which could potentially be mediated 
by supersymmetry, the physical value of $G$ would crucially depend 
on the precise value of the cutoff.
This might not seem appealing. Also, the 
approach outlined in Refs.~\cite{Sa2000grg,Vi2002} does not 
automatically lead to a formula expressing $G$ in terms of $\alpha_{\rm QED}$,
or vice versa. If the 
quantum fluctuation ansatz formulated in Refs.~\cite{Sa2000grg,Vi2002}
is to yield a connection of $G$ and $\alpha_{\rm QED}$,
then it will be necessary to study the quantum fluctuations 
of all quantized fields in the Universe, including the electrically 
charged fermion fields (see also Sec.~\ref{sec5C} below).
The quantum fluctuation-inspired approaches 
are important because, if containing elements of
truth, they will most likely imply at least a partial 
dismissal of the quantum gravity program, 
as gravitation will most likely 
no longer be seen as a fundamental interaction.

Finally, let us briefly mention that 
recently~\cite{On2013,On2013epl}, a model has been put 
forward which modifies gravity, at short distance 
scales, into a theory which incorporates 
quantum fluctuations of the weak gauge bosons.
The result is a modified gravitational force law,
of the form
\begin{equation}
\label{Veff}
V_{\rm eff}(\vec r) = - \frac{G_N \, m_1 \, m_2}{r} \,
\left[ 1 + \left( \frac{{\tilde G}_N}{G_N} - 1 \right) \,
\exp\left(-\frac{r}{{\tilde \Lambda}_p} \right)\right] \,,
\qquad
{\tilde G}_N = \frac{1}{\sqrt{2}} \,
\left( \frac{c}{\hbar} \right)^2  \, G_F \,,
\qquad
{\tilde \Lambda}_p = \sqrt{ \frac{2 \hbar {\tilde G}_N }{c^3} } \,.
\end{equation}
Here, ${\tilde G}_N$ is a modified gravitational 
constant into which $G_N$, the Newtonian gravitational constant,
is assumed to ``morph'' at short distances ($G_F$ is the 
Fermi coupling constant),
while ${\tilde \Lambda}_p$ is the ``Planck length'' corresponding to the 
``short-distance-version'' of gravity.
The model formulated in Refs.~\cite{On2013,On2013epl}
might explain the proton radius puzzle~\cite{PoEtAl2010,AnEtAl2013}
connected with the Lamb shift in muonic hydrogen, 
while its generalization to time-like momentum transfer might 
lead to additional corrections to the 
electron and muon $g$ factors~\cite{Je2011aop2},
which remain to be explored.
Also, a unification of the gravito-weak model formulated in 
Refs.~\cite{On2013,On2013epl} with electromagnetism is likely 
to yield correction terms to Eq.~\eqref{Veff}.
Still, the unification ansatz formulated in Refs.~\cite{On2013,On2013epl} is 
interesting because it highlights the need to 
search for clues toward unifications of gravity 
with other fundamental interactions.

%
%
\section{Conjectures Based on Quantum Field Theory}
\label{sec5}

%
%
\subsection{Graviton--Photon Conversion and Gravito--Electromagnetic RG}
\label{sec5A}

Inspired by the works of Gertsenshtein~\cite{Ge1962}, and by Zel'dovich and
Novikov~\cite{ZeNo1983apsrmp}, let us consider the possibility of a connection of the
gravitational and electromagnetic interactions on the quantum level. In the
quantized gravitational interaction, with a metric $g_{\mu\nu}$ of the curved
space-time, the deviation $h_{\mu\nu}$ from the flat space-time metric
$\eta_{\mu\nu} = {\rm diag}(1,-1,-1,-1)$ is quantized.  The quantized
$h_{\mu\nu}$ field operator (graviton) [see Eq.~(5) of Ref.~\cite{AbHa1986}] is
proportional to $\sqrt{G}$ where $G$ is Newton's gravitational constant.  The
investigation of quantized gravity and relevant Ward
identities~\cite{BaSh2005,BaNuScVi2007,BaSh2007,BaCoDaSc2012} as well as the
influence of gravitational interactions on the RG evolution of other gauge
couplings~\cite{RoWi2006} is a matter of ongoing investigations.
These combined gravito-electromagnetic models significantly enhance the scope
of nonlinear field-theoretical extensions of Maxwell theory 
alone~\cite{BoIn1934,In1936}.

In particular, we consider the tree-level amplitude contained in the coupling
$h_{\mu\nu} \, T^{\mu\nu}$ of the graviton $h_{\mu\nu}$ to the energy-momentum
tensor $T^{\mu\nu}$ of the electromagnetic field,
\begin{equation}
T^{\mu\nu} = F^{\mu\alpha} \, {F^\nu}_\alpha - 
\frac14 \, F_{\alpha\beta} \, F^{\alpha\beta} g^{\mu\nu} \,.
\end{equation}
Taking $F^{\mu\nu} = F_{\rm ext}^{\mu\nu} + f^{\mu\nu}$,
where $F_{\rm ext}^{\mu\nu}$ is the external field-strength 
tensor and $f^{\mu\nu}$ the photon field, the term 
$h_{\mu\nu} \, T^{\mu\nu}$ yields the trilinear form~\cite{Ge1962,ZeNo1983apsrmp,BaSh2007}
\begin{equation}
h_{\mu\nu} \, \left( F_{\rm ext}^{\mu\alpha} \, {f^\nu}_\alpha + 
{f^\mu}_\alpha \, F_{\rm ext}^{\nu\alpha} \right)
- \frac12 \, {g^\mu}_\mu \, F_{\rm ext}^{\alpha\beta} \, f_{\alpha\beta} \,,
\end{equation}
which mixes the graviton $h_{\mu\nu}$ and the photon~${f^\mu}_\alpha$. Indeed,
photon-graviton mixing near a pulsar 
has been considered in Ref.~\cite{RaSt1988}.
The possibility of converting photons into gravitons
may hint at a deeper connection of the 
two interactions, even in the low-energy domain.
In any case, the existence of graviton-photon mixing 
implies that the running of the QED and gravitational
coupling constants cannot be treated independently
of each other~\cite{RoWi2006}.

Indeed, the RG evolution of the fine-structure 
constant naturally leads to a conjecture involving the 
Planck and electron mass scales.
Generally, it is assumed that the coupling 
constants of physics unify at a length scale commensurate with the 
Planck scale, 
\begin{equation}
\ell_P = \left( \frac{\hbar G}{c^3}\right)^{1/2} = 
1.616 \times 10^{-25} \, {\rm m} \,,
\end{equation}
where we restore the factors of $\hbar$ and $c$.
The basic paradigm~\cite{BiBr2003} is that the 
coupling strengths of gravity and the electroweak model 
increase with the momentum scale, whereas the coupling
constant of quantum chromodynamics decreases. At $\ell_P$, they
are generally assumed to unify at a value of order unity 
(see, e.g., Ref.~\cite{BiBr2003}).
If we assume $\alpha_{\rm QED}$ to be of order unity 
at the Planck scale, and consider that the natural 
length scale of quantum electrodynamics is the mass
scale of the lightest fermion, the electron,
and furthermore conjecture that the logarithmic one-loop
running of the QED coupling approximates the 
full RG evolution sufficiently well, then 
it is natural to conjecture that 
\begin{equation}
\alpha_{\rm QED} 
\sim \left[ \ln \left( \frac{\Lambda}{m_e}\right)\right]^{-1}
\sim \left[ \ln \left( \frac{\lambdabar_e}{\ell_P}\right)\right]^{-1} \,,
\end{equation}
where $\lambdabar_e$ is the electron's Compton wavelength,
and $\Lambda$ is the cutoff (this formula also illustrates that 
the QED coupling constant goes to zero as the cutoff goes to 
infinity~\cite{ZJ2002}. Indeed, in Refs.~\cite{La1955apsrmp,TeEtAl1976,TeEtAl1977},
different approaches have led to relations 
similar to Eq.~(17) of Ref.~\cite{TeEtAl1977},
which we quote here as
\begin{equation}
\label{landau}
\alpha_{\rm QED} = 
\frac{3 \pi}{ (\sum Q^2)}
\left[ \ln \left( \frac{4 \pi}{\kappa_0 \, N_0 \, G \, m_e^2} \right)\right]^{-1} \,,
\end{equation}
where $\kappa_0 = 5/9$ or $2/3$, depending on whether a smooth,
Lorentz invariant or straight cutoff is used in the RG~\cite{TeEtAl1977}.
In Eq.~\eqref{landau}, the number $N_0$ is related to the number of 
active fermions in a Weinberg--Salam
multiplet. In particular, according to Ref.~\cite{TeEtAl1977},
we have $N_0 = \tfrac{15}{2} \, N$ and $\sum Q^2 = \tfrac83 \, N$
for $N$ multiplets (generations).
If the numerical prefactor in Eq.~\eqref{landau} is exact, then
a brief numerical calculation suggests 
there should exist five or six additional 
heavy charged leptons, neutrinos, and quark multiplets,
as noted by the authors of Ref.~\cite{TeEtAl1977}.
The extra generations have not been experimentally confirmed up to now.
One should also note that the precise evaluation of the 
prefactor in Eq.~\eqref{landau} may have to be revisited
and may depend on the details of the unification model used.

%
%
\subsection{Kaluza--Klein Theories and the Fine--Structure Constant}
\label{sec5B}

It is well known that Kaluza--Klein theories~\cite{Ka1921apsrmp,Kl1926}
represent one of the first attempts to unify gravity and electromagnetism.
It is less well appreciated that these theories also predict 
a direct proportionality of the gravitational and electromagnetic 
coupling constants. In order to illustrate this point,
let us start by recalling that the compactification of the 
fifth dimension in Kaluza--Klein theories
leads to a natural explanation of charge quantization,
while relating the fine-structure constant to 
a quantity which can either be interpreted as the vacuum expectation
value of a scalar field, or, to a physical constant 
which relates the fifth components of the metric 
to the $(4 \times 4)$-space-time components.
The covariant coupling in QED is of course given by the 
covariant derivative, $p_\mu \to p_\mu - e \, A_\mu $,
or formulated differently, by the replacement
$\partial_\mu \to \nabla_\mu = \partial_\mu + \ii \, e \, A_\mu$.
Let $x = (x^0, x^1, x^2, x^3)$ denote a four-vector,
while $y$ adds the fifth dimension.
Let $A, B$, denote indices in the extended space,
i.e., including the extra dimensions.
We start from Eq.~(22) of Ref.~\cite{OvWe1998},
\begin{align}
S_{\hat\psi} =& \;
- \int \dd^4 x \, \dd y \,
\sqrt{- {\rm det}(\hat g) } \;\;
\partial^A \hat \psi(x,y) \;\; \partial_A \hat \psi(x,y) \,,
\nonumber\\[0.133ex]
\mu =& \; 0,1,2,3 \,,
\qquad
A = 0,1,2,3,4 \,,
\end{align}
with the metric given in Eq.~(16) of Ref.~\cite{OvWe1998},
\begin{equation}
\left( g_{AB} \right) = 
\phi^{-1/3} \, 
\left( \begin{array}{cc} 
g_{\alpha\beta} + \varkappa^2 \, \phi \, \hat A_\alpha \, \hat A_\beta & 
\quad
\varkappa \, \phi \, \hat A_\alpha \\[2ex]
\varkappa \, \phi \, \hat A_\alpha & 
\quad \phi \\
\end{array} \right) \,.
\end{equation}
Here, $\varkappa =  \sqrt{16 \pi \, \hat G}$.
The expansion into Fourier modes in the compactified 
dimension leads us to the formula
\begin{equation}
\hat \psi(x, y) = 
\sum_{n=-\infty}^{n=\infty}  
\psi^{(n)}(x) \; \ee^{\ii n y / r} \,,
\end{equation}
where the $\psi^{(n)}(x)$ denote massive scalar fields
coupled to the electromagnetic field,
while $r$ is the scale of the compactification 
of the fifth dimension. (We reserve the hat over a symbol
for the five-dimensional generalization of a 
quantity otherwise defined on four-dimensional space-time.)
The action is given as follows,
\begin{align}
S_{\hat\psi} =& \; - \left( \int \dd y \right) \, 
\sum_{n=-\infty}^{\infty} \, \int \dd^4 x \, \dd y \,
\sqrt{- {\rm det} g } \;\;
\nonumber\\[0.133ex] 
& \; \times \left\{ \left[ 
\left( (\ii \, \partial^\mu - \frac{n}{r} \, \varkappa \, \hat A^\mu \right) \,
\psi^{(n)}(x) \right] \,
\left[ \left( \ii \, \partial_\mu - \frac{n}{r} \, \varkappa \, \hat A_\mu \right) \,
\psi^{(n)}(x) \right]
- \frac{n^2}{\phi \, r^2} \, 
\left( \hat \psi^{(n)}(x) \right)^2 \right\} \,,
\end{align}
where $g$ is the four-dimensional restriction of $\hat g$.
We now renormalize the four-vector potential 
and define the four-dimensional gravitational coupling constant
as follows,
\begin{equation}
\hat A_\alpha  = \frac{A_\alpha}{ (\phi \int \dd y)^{1/2}} \,,
\qquad
\qquad
G = \frac{\hat G}{\int \dd y} \,.
\end{equation}
The scaling of the four-vector potential is necessary in 
order to eliminate an otherwise disturbing factor $\phi$ in front of the 
gauge boson term in the action functional
(which we do not explicitly write out, see Ref.~\cite{OvWe1998}
for details).
Otherwise, this term blows up when we choose $\phi$ to be large.
The sum over spinless fields, coupled to the electromagnetic 
field, is given as follows,
\begin{align}
S_{\hat\psi} =& \; - \left( \int \dd y \right) 
\sum_{n=-\infty}^{\infty} \, \int \dd^4 x \; \dd y \,
\sqrt{- {\rm det} \, g } \;\;
\left\{ \left[ 
\left( (\ii \, \partial^\mu - \frac{n \, \varkappa}{r (\phi \int \dd y)^{1/2}} \, 
A^\mu \right) \,
\psi^{(n)}(x) \right] \;
\right.
\nonumber\\[0.133ex]
& \; 
\times 
\left.
\left[ \left( \ii \, \partial_\mu - \frac{n \, \varkappa}{r (\phi \int \dd y)^{1/2}} \,
A_\mu \right) \psi^{(n)}(x) \right]
- \frac{n^2}{\phi \, r^2} \, \left( \psi^{(n)}(x) \right)^2 \right\} \,.
\end{align}
Charge is found to be naturally quantized according to the formula
\begin{equation}
q_n = \frac{n \, \varkappa}{r \, (\phi \int \dd y)^{1/2}}
= \frac{n \, \sqrt{16 \, \pi}}{r \, (\phi \int \dd y)^{1/2}} \, \sqrt{\hat G}
= \frac{n \, \sqrt{16 \, \pi}}{r \, \sqrt{\phi}} \, \sqrt{G} \,.
\end{equation}
Setting $n=1$, one obtains
\begin{equation}
\label{KKALPHA}
\alpha_{\rm QED} = \frac{e^2}{4 \pi} = \frac{q^2_1}{4 \pi} 
= \frac{16 \, \pi}{4 \pi (r \, \sqrt{\phi})^2} \, G
= \frac{4}{(r \, \sqrt{\phi})^2} \, G \,,
\end{equation}
suggesting a proportionally of the 
electromagnetic and gravitational fine-structure
constants. If the field $\phi$ varies in space and/or time,
this could otherwise explain a time variation of the 
fundamental constants.
The main problem of the Kaluza--Klein formalism 
is due to the very large
generated mass parameters, $m^2 = n^2/(\phi \, r^2)$ which are
obtained when a realistic physical value for the 
quantity $(r \, \sqrt{\phi})$,
determined from Eq.~\eqref{KKALPHA},
is inserted into the formula $m^2 = n^2/(\phi \, r^2)$.
Another problem is related to the instability of the radius $r$ of the compact
dimension against small perturbations.
None of these problems have been satisfactorily addressed
in the literature up to this point, while the 
general idea of Kaluza and Klein has inspired many 
technical developments in field theory over the last decades.

%
%
\subsection{String Theory and Connections of Gravity and Electromagnetism}
\label{sec5C}

String theories, and superstring theories contain both gravitational and gauge
interactions~\cite{Po1998vol1,Po1998vol2}. In principle, they thus offer a
possible way to unify the gauge field and gravity.  Gravitons and gauge
particles are assumed to correspond to massless states of closed and open
strings. Relations between closed and open strings then imply the relations
between gravitational interactions and gauge fields.  In the consideration of
strings on a curved space-time background~\cite{Po1998vol1}, it is almost
universally assumed [see Eq.~(3.2.7) of Ref.~\cite{Po1998vol1}] that a
relationship of the form
\begin{equation}
\label{STRING}
g_o^2 \sim g_c \sim \exp(\lambda)
\end{equation}
holds, where $g_o$ is the gauge coupling, $g_c$ is the gravitational coupling,
and $\lambda$ parameterizes the string interaction.  According to the text
following Eq.~(3.7.17) of Ref.~\cite{Po1998vol1}, the value of $\lambda$
determines the coupling strength between strings, but this does not necessarily
imply that string theories with different values of this parameter describe
different physics; namely, $\lambda$ defines a basic unit of length, and can be
absorbed in a redefinition of the coordinates $X^\mu$ of the string world
sheet. Specialized to quantum chromodynamics,
analogous relations are also known as the Kawai--Lewellen--Tye 
relations~\cite{KaLeTy1986}.

Now, combining Eqs.~(3.7.26) and Eq.~(6.6.18)
of Ref.~\cite{Po1998vol1}, we have
\begin{equation}
\kappa = (8 \pi G)^{1/2} = 2 \pi \, g_c \,,
\qquad
g^2_c = \frac{\kappa^2}{4 \pi^2} =
\frac{2}{\pi} \, G \,.
\end{equation}
According the Eq.~(3.6.26) of Ref.~\cite{Po1998vol1},
the photon vertex operator is given as
\begin{equation}
-\ii \, \frac{g_o}{(2 \alpha')^{1/2}} \, e_\mu \, 
\int_{\partial M} \dd s \, 
\left[ {\dot X}^\mu \, \exp(\ii k \cdot X) \right]_{\Gamma} 
\end{equation}
where $\alpha'$ is a numerical parameter 
of order unity [$\alpha' = 2$ for the closed string, while
$\alpha' = \tfrac12$ for the open string],
$k$ is the exchange four-momentum,
$e_\mu $ the polarization vector, and the interaction
is evaluated on the string world-sheet $(\Gamma, \partial M)$.
Furthermore, according to Eq.~(6.6.23) of Ref.~\cite{Po1998vol1},
the four-point open (gauge coupling) $A_o$ and 
closed (gravitational) string amplitudes $A_c$ are related
as follows,
\begin{equation}
\label{AMPLITUDES}
g_o^4 \, A_c(s,t,u, \alpha', g_c) = 
g_c^2 \, \pi \, \ii \, \alpha' \, 
\sin[ \pi \, \alpha_o(t) ] \,
A_o(s,t,\tfrac14 \, \alpha', g_o) \,
A_o(t,u,\tfrac14 \, \alpha', g_o)^* \,.
\end{equation}
where $\alpha_o(t) = 1 + \alpha' \, x$,
and the $s$, $t$ and $u$ are the Mandelstam variables.
Both the basic equation~\eqref{STRING} and
the more detailed formulation given in Eq.~\eqref{AMPLITUDES} 
suggest a proportionality of the form,
inspired by string theory,
\begin{equation}
\label{stringconj}
\alpha_{\rm QED} \propto \sqrt{G} \,,
\end{equation}
when expressed in terms of the physical couplings 
of QED and gravity. We recall Eq.~\eqref{alphaG}, which implies that 
the gravitational fine-structure constant of electron and proton
reads as
\begin{equation}
\alpha_G = \frac{G \, m_e \, m_p}{\hbar c} \,.
\end{equation}
Based on the physical values of $\alpha_{\rm QED}$ and 
$\alpha_G$, and the inspiration from string theory~\cite{Po1998vol1,Po1998vol2},
one may attempt to express the proportionality 
factor in Eq.~\eqref{stringconj} in a simple form.
Based on the  coincidence
\begin{equation}
\frac{\alpha_{\rm QED}}{\sqrt{ \alpha_G }} \approx 4.07 \times 10^{18} \,,
\qquad
\qquad
\exp\left( \sqrt{\frac{m_p}{m_e}} \right) \approx 4.07\times 10^{18} \,,
\end{equation}
one may investigate the following relationship,
which is numerically fulfilled to relatively good accuracy,
\begin{equation}
\label{rel}
\frac{1}{\sqrt{\alpha_G}} \, 
\exp\left( -\sqrt{\frac{m_p}{m_e}} \right) =
\left( \frac{G \, m_e \, m_p}{\hbar c} \right)^{-1/2} \;
\exp\left( -\sqrt{\frac{m_p}{m_e}} \right) =
136.976(8) \approx \frac{1}{\alpha_{\rm QED}} \,.
\end{equation}
This relation is consistent with the ``string-inspired conjecture''
$\alpha_{\rm QED} \propto \sqrt{G}$
and identifies the proportionality factor as 
being approximately given by the expression
$[\hbar \, c/(m_e \, m_p)]^{1/2} \, \exp[ (m_p/m_e)^{1/2} ]$.
In Eq.~\eqref{rel}, the latest CODATA value~\cite{MoTaNe2012} for the gravitational 
constant has been used, $G = 6.67384(80) \times 10^{-11} 
\, {\rm N} \, {\rm m}^2 \, {\rm kg}^{-2}$.
Its uncertainty dominates the uncertainty of the numerical value of the 
expression on the left-hand side of the relation~\eqref{rel}.
One may observe that recent determinations of Newton's gravitational 
constant are not all in mutual agreement~\cite{Gi1997};
the scatter of the experimental data has not 
been resolved~\cite{Mo2014}.
One should add that observations related to Eq.~\eqref{rel} 
had previously been made in Ref.~\cite{Br1992ieee} and
interpreted as a means of determining $G$, not $\alpha_{\rm QED}$.

The square root of the string-inspired on the right-hand
side of Eq.~\eqref{stringconj} otherwise finds a physical
interpretation as follows.
Namely, according to Ref.~\cite{JeNo2013pra},
there exists a relativistic correction 
term in the gravitational bound-state problem
which can naturally be identified as the gravitational 
quiver motion (zitterbewegung) term and which is 
proportional to 
\begin{equation}
H_Z \propto \frac{\hbar^2 \, G \, m_p}{c^2 \, m_e} \, 
\delta^{(3)}(\vec r) 
\equiv \hbar \, c\, r_Z^2 \,
\delta^{(3)}(\vec r)  \,,
\end{equation}
where $r_Z$ is the radius of the quiver motion (zitterbewegung) 
of the gravitationally bound electron.
The identification of the Darwin term with the 
quiver motion is discussed on p.~71 in Sec.~2.2.4 of Ref.~\cite{ItZu1980}.
Solving for $r_Z^2$, we obtain
\begin{equation}
r_Z^2 = \frac{\hbar \, G \, m_p}{c^3 \, m_e} \, 
= \frac{\hbar \, G}{c^3} \, \frac{m_p}{m_e} 
= \ell_P^2 \, \frac{m_p}{m_e} \,,
\end{equation}
where $\ell_P = ( \hbar G / c^3 )^{1/2}$ is the Planck 
length. The argument of the exponential factor in Eq.~\eqref{rel} is 
thus identified as the ratio 
of the gravitational zitterbewegung term to the Planck length,
\begin{equation}
\left( \frac{m_p}{m_e} \right)^{1/2} = \frac{r_Z}{\ell_P} \,.
\end{equation}
It appears that the square root of the mass ratio
of the proton and electron naturally 
occurs in the gravitational bound-state problem.

Let us also consider the following scaling transformation
(a ``global dilaton'') of the fields,
\begin{subequations}
\label{conf2}
\begin{equation}
A^\mu \to \lambda \, A^\mu \,,
\qquad
A_\mu \to \lambda \, A_\mu \,,
\qquad
\psi \to \lambda \, \psi \,,
\end{equation}
combined with the following transformation of the coordinates,
\begin{equation}
x^\mu \to \lambda^{-{1/2}} \, x^\mu \,,
\qquad
x_\mu \to \lambda^{1/2} \, x_\mu \,,
\qquad
g_{\mu\nu} \to \lambda g_{\mu\nu} \,,
\qquad
g^{\mu\nu} \to \lambda^{-1} \, g^{\mu\nu} \,,
\end{equation}
\end{subequations}
where the transformation of the 
metric entails the following 
transformation of the curved-space 
Dirac matrix, 
$ \overline \gamma^\mu \to \lambda^{-1/2} \, \overline \gamma^\mu$.
This transformation modifies the Einstein--Maxwell Lagrangian,
originally given as
\begin{align}
\label{Sorig}
S = & \; \int \dd^4 x \, \sqrt{- {\rm det} \, {g}} \,
\left\{ \frac{R}{16 \, \pi \, G} - \frac14 \, F^{\mu\nu} \, F_{\mu\nu} 
+ {\overline \psi}(x) \; \left[ \ii {\overline \gamma}^\mu \;
\left( \nabla_\mu - e \, A_\mu \right) - m \right] \, \psi(x) \right\}  \,,
\end{align}
into
\begin{align}
S' =& \; \int \frac{\dd^4 x}{\lambda^2} \, \sqrt{- 
{\color{black} \lambda^4} {\rm det} \, {g}} \,
\left\{ \frac{R}{16 \, \pi \, G} - \frac{ \lambda^2}{4} \, F^{\mu\nu} \, F_{\mu\nu} 
\right.
\nonumber\\[0.133ex]
& \; \left.
+ \lambda^{2} \; {\overline \psi}(x) \; 
\left[ \ii \lambda^{-1/2} \, {\overline \gamma}^\mu \;
\left( \lambda^{1/2} \nabla_\mu - e \, \lambda \, A_\mu \right) {\color{black} - m}
\right] \, \psi(x)
\right\} \,,
\end{align}
which can be rearranged into
\begin{equation}
\label{S2prime2}
S' = {\color{black} \lambda^2} \, \int \dd^4 x \, \sqrt{- {\rm det} \, {g}} \,
\left\{ \frac{R}{16 \, \pi \, G \, \lambda^2} - 
\frac{1}{4} \, F^{\mu\nu} \, F_{\mu\nu} 
+ {\overline \psi}(x) \; \left[ \ii {\overline \gamma}^\mu \;
\left( \nabla_\mu - e \, \lambda^{1/2} \, A_\mu \right) {\color{black} - m} \right] \, \psi(x)
\right\} \,.
\end{equation}
The coupling constants have transformed as follows,
\begin{equation}
G \to \lambda^2 \, G \,,
\qquad
e^2 \to \lambda \, e^2 \,.
\end{equation}
This scaling is consistent with the 
proportionality $\alpha_{\rm QED}^2 \propto 
e^4 \propto \lambda^2 \propto G \propto \alpha_G$,
and thus with the string-inspired conjecture 
given in Eq.~\eqref{rel}.
The conformal transformation~\eqref{conf2} leaves
the following action integrals
\begin{align}
\int \dd^4 x \, | \psi |^2 \to & \;
\left( \frac{1}{\lambda^{1/2}} \right)^4 \, 
\int \dd^4 x \, \left( \lambda \, | \psi | \right)^2 =
\int \dd^4 x \, | \psi |^2  \,,
\nonumber\\[0.133ex]
\int \dd^4 x \, F^{\mu\nu} \, F_{\mu\nu} \to & \;
\left( \frac{1}{\lambda^{1/2}} \right)^4 \, 
\int \dd^4 x \, \lambda^2 \, F^{\mu\nu} \, F_{\mu\nu}  \,,
\end{align}
invariant.
In order to arrive at a variational principle, 
we again isolate from Eq.~\eqref{S2prime2} the terms in the 
Lagrangian which depend on the scale parameter $\lambda$,
\begin{equation}
\label{L2}
S'' = \frac{S'}{\lambda^2} = 
\int \dd^4 x \, \sqrt{- {\rm det} \, {g}} \, \calL' \,,
\qquad
\calL' \to \frac{R}{16 \, \pi \, G \, \lambda^2} -
e \,\lambda^{1/2} \, {\overline \psi}(x) \; A_\mu \; \psi(x) \,.
\end{equation}
Most particles in the Universe are in bound states,
both gravitationally as well as electromagnetically.
Integrating the Lagrangian density over all space,
one obtains the Lagrangian of the Universe.
As matter is found in the bound states of atoms,
we have as an order-of-magnitude estimate
\begin{equation}
\label{E0}
-\lambda^{1/2} \int \dd^3 x \; \left< e \, \bar\psi(x) \; A_\mu \;  \psi(x) \right> 
\sim \lambda^{1/2}  \, E_0 \sim \lambda^{1/2}  \, 27.2 \, {\rm eV} \,,
\end{equation}
where $E_0$ is a typical binding energy in an atom,
which is distributed over the electrons and protons,
measured in terms of the Hartree energy of $27.2 \, {\rm eV}$,
and $\psi$ can be the wave function of a proton or an electron.

The gravitational contribution per proton can be
estimated as the contribution to the 
curvature integral of the Schwarzschild metric 
for an average star,
\begin{equation}
\label{EG}
\frac{1}{\lambda^2} \,
\int \dd^3 x \; \frac{R}{16 \, \pi \, G} 
\sim \frac{1}{\lambda^2} \,
\left( G \, \frac{m_p \, M_{\rm Sun}}{r_{\rm Sun}} \right)
= \frac{1}{\lambda^2} \, E_G 
\sim \frac{1}{\lambda^2} \, 1991.3 \, {\rm eV} \,.
\end{equation}
We here use the radius $r_{\rm Sun}$ and its mass
$m_{\rm Sun}$ in our solar
system as a measure of an average particle bound to a
typical star in the Universe.
If $\lambda$ acts as a variational parameter,
then the variational condition reads as 
\begin{equation}
\label{var2}
\frac{\partial}{\partial \lambda} \,
\left( 
\frac{1}{\lambda^2} \, E_G 
+ \lambda^{1/2} \, E_0 \right) \mathop{=}^{\mbox{!}} 0 \,,
\qquad
\lambda = 2^{4/5} \, \left( \frac{E_G}{E_0} \right)^{1/3} \,.
\end{equation}
With our order-of-magnitude estimates~\eqref{E0} and~\eqref{EG},
we obtain $\lambda \approx 9.70$ for the current Universe,
which differs from unity by less than one order of magnitude.
A more precise variational calculation remains elusive to 
this date and requires a better understanding of 
the mass distribution in the Universe (``dark matter'').

%
%
\section{Conclusions}
\label{sec6}

We have discussed attempts to find approximate formulas for the fine-structure
constant, based on purely algebraic relations (see Sec.~\ref{sec2}).  The most
promising ansatz in this direction has been a hypothetical zero of the QED
$\beta$ function.  However, it appears that nature does not do us the favor of
providing a fixed point for the RG evolution of the QED coupling in the
low-energy regime which could form a suitable candidate equation for the
determination of $\alpha_{\rm QED} \approx 1/137.036$ (see Sec.~\ref{sec2A}).
Group-theoretical approaches discussed in Sec.~\ref{sec2B} are very rigid in
predicting fixed numerical values for $\alpha$ which therefore cannot  receive
any further quantum field-theoretical loop corrections, but so far, no
compelling predictions have been obtained for manifestly interacting particles.

Arguments based on classical physics (see Sec.~\ref{sec3}) meet considerable
difficulty when confronted with the additional wisdom obtained from modern
particle physics experiments, which challenge some of the concepts on which the
original conjectures were based. Theories relating 
classical and quantum fluctuations (e.g., of electromagnetic origin)
to gravitational interactions (see Sec.~\ref{sec4} have yet 
to produce a viable conjecture for a formula that would 
relate the fine-structure constant to other fundamental constants.

In turn,
conjectures inspired by quantum field-theory (see Sec.~\ref{sec5}) have been
strongly inspired by the graviton-photon conversion amplitude analyzed in
Refs.~\cite{Ge1962,ZeNo1983apsrmp} 
discussed in Sec.~\ref{sec5A}.  The renormalization-group
inspired formula~\eqref{landau}, studied in
Refs.~\cite{La1955apsrmp,TeEtAl1976,TeEtAl1977}, relates the QED coupling at
the Planck scale to the observed coupling parameter $\alpha_{\rm QED}$, which
is relevant to the low-energy (electron mass) scale.  We note that the RG
running of the gravitational and electromagnetic coupling constants are
intertwined~\cite{RoWi2006} so that Eq.~\eqref{landau}, even if it holds
approximately, will receive higher-order loop corrections from graviton-photon
conversion.  However, in terms of promising directions for future research, we
believe that it may be useful to put Eq.~\eqref{landau} into the context of
other conjectures which have been discussed in the literature.

In Sec.~\ref{sec5B}, we discuss Kaluza--Klein theories.  Often, it is not really
appreciated that Kaluza--Klein theories with a compactified fifth dimension
predict rather unambiguously that $\alpha_{\rm QED}$ and $G$ should be directly
proportional to each other [see Eq.~\eqref{KKALPHA}].  However, the mass
hierarchy problems connected with such models have never been resolved in the
literature, so that related conjectures should probably be taken with a grain
of salt.

String theory~\cite{Po1998vol1,Po1998vol2} predicts a different functional
relationship which implies that $\alpha_{\rm QED}$ should
be proportional to $\sqrt{G}$ [see Eq.~(3.2.7) of Ref.~\cite{Po1998vol1}].  In
turn, a dimensionless variant of the gravitational constant is given by the
electron-proton coupling strength given in Eq.~\eqref{alphaG}, which we refer
to as $\alpha_G$.  One may thus attempt to search for simple analytic
formulas which might describe the constant of proportionality in the predicted
relationship $\alpha_{\rm QED} \propto \sqrt{G}$.  Surprisingly, numerical
experimentation yields quite good agreement if one simply sets the
proportionality constant equal to $\exp(\sqrt{m_p/m_e})$ [see
Eq.~\eqref{stringconj}].  Most probably, this observation is a numerical
coincidence and does not allow us to gain any deeper insight; however, it may
be permissible to record the observation~\eqref{rel} as it may inspire
models which attempt to match a few of the parameters of string-theoretical,
and supersymmetric string-theory models with low-energy properties of the
Standard Model, notably, the electron and proton masses.

The future will tell if any of the conjectures survive the 
tests of scrutiny that will confront these with other developments
from the renormalization-group analysis of the intertwined gravitational
and electromagnetic interactions, or, more sophisticated string-theoretical
models, and experiments which may fix parameters of the models based on
independent relations.

%
%
\section*{Acknowledgements}

Helpful discussions with P. J. Mohr, M.~M.~Bush
and E. L\"{o}tstedt are gratefully acknowledged.
Work on this project has  been supported by the National Science Foundation 
(Grants PHY--1068547 and PHY--1403973).

\appendix

%
%
\section{Analogy of the Electrostatic and Gravitational Bound--State Problem}
\label{appa}

Recently, the gravitational bound-state problem has been investigated in a
relativistic quantum
framework~\cite{Je2013,JeNo2013pra,JeNo2014jpa,Je2014dirac,Je2014pra}.  Gravity
and QED are the only two long range interactions mediated by a massless gauge boson
(photon and graviton),
while gluons have never been observed as free particles 
due to their confinement into hadrons.
It is reasonable to search for  a connection of the
value of $\alpha_{\rm QED}$ to other physical constants, derived from the
low-energy regime alone, and the gravitational quantum bound-state problem
provides for an interesting starting point. 

The gravitational coupling constant for the interaction 
of electron and proton can be derived 
based on the Bohr--Sommerfeld quantization
condition alone. The centripetal force $F$ exerted on the electron 
in its circular orbit is as follows,
$F = m_e \, v^2/R$, where $R$ is the radius of the orbit.
On the circular orbit, the centripetal force is equal to the 
gravitational attractive force, and we have
\begin{equation}
F = \frac{m_e \, v^2}{R} = \frac{G m_e m_p}{R^2} \,,
\qquad
\frac12 \, m_e \, v^2 = \frac{G m_e m_p}{2 R} \,,
\end{equation}
where the latter equation is obtained from the former
via multiplication by $R/2$ (the 
modulus of the electron velocity in the circular orbit is $v$).
The virial theorem states that 
\begin{equation} 
\label{virial}
E = \frac12 \, m_e \, v^2 - \frac{G m_e m_p}{R} = - \frac{G m_e m_p}{2 R} \,,
\end{equation}
i.e.~the potential energy is twice as large as the kinetic 
energy on the circular orbit, and has the opposite sign.
Let us now implement the Bohr--quantization condition
in its most basic form,
\begin{equation} 
\label{quant}
\int p \, \dd q = n \, h \,,
\qquad
m_e \, v \, (2\pi R) = n \, h \,,
\qquad
m_e \, v \, R = n \, \hbar \,.
\end{equation} 
Let us square the latter equation,
\begin{equation} 
\frac{(n \, \hbar)^2}{2 m_e} =
\frac{(m_e \, v \, R )^2}{2 m_e} =
\frac{m_e v^2}{2} \, \, R^2 =
\frac{G m_e m_p}{2 R} \, R^2 \,,
\end{equation} 
so that, on the quantized orbits, we have
\begin{equation} 
(n \, \hbar)^2 = G m_e^2 m_p \, R \,.
\end{equation}
We plug this into the virial theorem
and obtain the result
\begin{equation} 
\label{schr}
E_n = - \frac12 \, 
\left( \frac{G m_e m_p}{\hbar\, c} \right)^2 \, \frac{m_e \, c^2}{n^2}
= - \frac12 \, \alpha_G^2 \, \frac{m_e \, c^2}{n^2} \,,
\end{equation} 
where we use the ``gravitational fine-structure constant'' as~\cite{Je2014dirac}
\begin{equation}
\label{alphaG}
\alpha_G = \frac{G \, m_e \, m_p}{\hbar c} \,.
\end{equation}
The numerical value is about $3.21 \times 10^{-42}$.
The gravitational Schr\"{o}dinger formula~\eqref{schr}
is based on the quantization of the orbit according to 
Eq.~\eqref{quant}. By way of comparison,
the Schr\"{o}dinger formula for the 
electromagnetically bound atomic hydrogen energy levels is 
\begin{equation} 
\label{schr_qed}
E_n = - \frac12 \, \alpha_{\rm QED}^2 \, \frac{m_e c^2}{n^2} \,.
\end{equation} 
Equation~\eqref{schr_qed}
is obtained from Eq.~\eqref{schr} by the simple substitution 
$\alpha_G \to \alpha_{\rm QED}$.
Based on an analysis of the Dirac--Schwarzschild
central-field problem in general relativity~\cite{Je2013,JeNo2013pra,Je2014dirac},
it has recently been verified that $\alpha_G$
is the analogue of the QED coupling constant for 
gravitational phenomena, on the basis of a
calculation of the relativistic corrections for the 
Dirac-Schwarzschild central-field problem~\cite{JeNo2013pra,Je2014dirac}.


\vspace*{1cm}

\begin{thebibliography}{10}

\bibitem{Di1938}
P.~A.~M. Dirac, Proc. Roy. Soc. London, Ser. A {\bf 167},  148  (1938).

\bibitem{FiEtAl2004}
M. Fischer, N. Kolachevsky, M. Zimmermann, R. Holzwarth, T. Udem, T.~W.
  H\"{a}nsch, M. Abgrall, J. Gr\"unert, I. Maksimovic, S. Bize, H. Marion, F.
  Pereira Dos~Santos, P. Lemonde, G. Santarelli, P. Laurent, A. Clairon, C.
  Salomon, M. Haas, U.~D. Jentschura, and C.~H. Keitel, Phys. Rev. Lett. {\bf
  92},  230802  (2004).

\bibitem{Wy1969}
A. Wyler, C. R. Acad. Sci. Paris Ser. A--B {\bf 269},  A743  (1969).

\bibitem{Wy1971}
A. Wyler, C. R. Acad. Sci. Paris Ser. A {\bf 271},  186  (1971).

\bibitem{GMLo1954}
M. Gell-Mann and F.~E. Low, Phys. Rev. {\bf 95},  1300  (1954).

\bibitem{BaJo1969}
M. Baker and K. Johnson, Phys. Rev. {\bf 183},  1292  (1969).

\bibitem{AkBe1969}
A.~I. Akhiezer and V.~B. Berestetskii, {\em \relax{Quantum Electrodynamics}}
  (Nauka, Moscow, 1969).

\bibitem{Ad1972}
S.~L. Adler, Phys. Rev. D {\bf 5},  3021  (1972).

\bibitem{Ma1984rg}
E.~B. Manoukian, Fortschr. Physik {\bf 32},  315  (1984).

\bibitem{ZJ2007}
J. Zinn-Justin, {\em \relax{Phase Transitions and Renormalisation Group}}
  (Oxford University Press, New York, 2007).

\bibitem{WeFlChDrBa1999}
J.~K. Webb, V.~V. Flambaum, C.~W. Churchill, M.~J. Drinkwater, and J.~D.
  Barrow, Phys. Rev. Lett. {\bf 82},  884  (1999).

\bibitem{DzFlWe1999}
V.~A. Dzuba, V.~V. Flambaum, and J.~K. Webb, Phys. Rev. Lett. {\bf 82},  888
  (1999).

\bibitem{WeEtAl2001}
J.~K. Webb, M.~T. Murphy, V.~V. Flambaum, V.~A. Dzuba, J.~D. Barrow, C.~W.
  Churchill, J.~X. Prochaska, and A.~M. Wolfe, Phys. Rev. Lett. {\bf 87},
  091301  (2001).

\bibitem{MuWeFl2003}
M.~T. Murphy, J.~K. Webb, and V.~V. Flambaum, Mon. Not. Roy. Astron. Soc. {\bf
  345},  609  (2003).

\bibitem{WeEtAl2011}
J.~K. Webb, J.~A. King, M.~T. Murphy, V.~V. Flambaum, R.~F. Carswell, and M.~B.
  Bainbridge, Phys. Rev. Lett. {\bf 107},  191101  (2011).

\bibitem{Ka1921apsrmp}
T. Kaluza, Preussische Akademie der Wissenschaften (Berlin), Sitzungsberichte,
  966--972, 1921.

\bibitem{Kl1926}
O. Klein, Z. Phys. A {\bf 37},  895  (1926).

\bibitem{OvWe1998}
J.~M. Overduin and P.~S. Wesson, Phys. Rep. {\bf 283},  303  (1997).

\bibitem{Po1998vol1}
J. Polchinski, {\em \relax{String Theory (Volume 1): An Introduction to the
  Bosonic String}} (Cambridge University Press, Cambridge, UK, 1989).

\bibitem{Po1998vol2}
J. Polchinski, {\em \relax{String Theory (Volume 2): Superstring Theory and
  Beyond}} (Cambridge University Press, Cambridge, UK, 1989).

\bibitem{dRRo1974}
E. de~Rafael and J.~L. Rosner, Ann. Phys. (N.Y.) {\bf 82},  369  (1974).

\bibitem{ItZu1980}
C. Itzykson and J.~B. Zuber, {\em \relax{Quantum Field Theory}} (McGraw-Hill,
  New York, 1980).

\bibitem{GoKaLaSu1991}
S.~G. Gorishny, A.~L. Kataev, S.~A. Larin, and L.~R. Surguladze, Phys. Lett. B
  {\bf 256},  81  (1991).

\bibitem{Ca1970}
J. Callan, Phys. Rev. D {\bf 2},  1541  (1970).

\bibitem{Sy1970}
K. Symanzik, Commun. Math. Phys. {\bf 18},  227  (1970).

\bibitem{ZJ2002}
J. Zinn-Justin, {\em \relax{Quantum Field Theory and Critical Phenomena}}, 4th
  ed. (Oxford University Press, Oxford, 2002).

\bibitem{DuGiSc2002}
G.~V. Dunne, H. Gies, and C. Schubert, J. High Energy Phys. {\bf 11},  032
  (2002).

\bibitem{Ad1972fermilab}
S.~L. Adler, {\em Theories of the Fine--Structure Constant $\alpha$}, Fermilab
  publication FERMILAB--PUB--72/059--T, 1972.

\bibitem{LaAbKh1954}
L.~D. Landau, A.~A. Abrikosov, and I.~M. Khalatnikov, Dokl. Akad. Nauk SSSR
  {\bf 95},  1177  (1954).

\bibitem{KiKoLo2002}
S. Kim, J.~B. Kogut, and M.-P. Lombardo, Phys. Rev. D {\bf 65},  054015
  (2002).

\bibitem{GoEtAl1997}
M. G\"{o}ckeler, R. Horsley, V. Linke, P. Rakow, G. Schierholz, and H.
  St\"{u}ben, Phys. Rev. Lett. {\bf 80},  4119  (1997).

\bibitem{GiJa2004}
H. Gies and J. Jaeckel, Phys. Rev. Lett. {\bf 93},  110405  (2004).

\bibitem{Su2001}
I.~M. Suslov, Pis'ma v ZhETF {\bf 74},  211  (2001), [JETP Lett.~{\bf 74}, 191
  (2001)].

\bibitem{Su2009}
I.~M. Suslov, Zh. \'{E}ksp. Teor. Fiz. {\bf 135},  1129  (2009), [JETP {\bf
  108}, 980 (2009)].

\bibitem{BiBr2003}
M. Binger and S.~J. Brodsky, Phys. Rev. D {\bf 69},  095007  (2003).

\bibitem{JaMa2012}
A. Jakov\'{a}c and P. Mati, Phys. Rev. D {\bf 85},  085006  (2012).

\bibitem{JaMa2013}
A. Jakov\'{a}c and P. Mati, Phys. Rev. D {\bf 87},  125007  (2013).

\bibitem{JaMa2014}
A. Jakov\'{a}c and P. Mati, Phys. Rev. D {\bf 90},  045038  (2014).

\bibitem{Dy1952}
F.~J. Dyson, Phys. Rev. {\bf 85},  631  (1952).

\bibitem{GuZJ1998}
R. Guida and J. Zinn-Justin, J. Phys. A {\bf 31},  8103  (1998).

\bibitem{ScWi2012}
E. Sch\"{o}nfeldt and P. Wilde, Progress in Physics {\bf 1},  3  (2012).

\bibitem{Da2011}
G. Dattoli, {\em The fine structure constant and numerical alchemy}, e-print
  arXiv:1109.1711 [physics.gen-ph], 2011.

\bibitem{HuMAGICapsrmp}
L.~K. Hua, in {\em American Mathematical Society Translations (American
  Mathematical Society, Providence, R. I.)}, vol.~6, pp.~6--8, pp. 48, 97, and
  vol.~32, p.~195, Eqs. (11) and (65), and p.~217, 1963.

\bibitem{Pe1977}
R.~L. Pease, Int. J. Theor. Phys. {\bf 16},  405  (1977).

\bibitem{Ro1971}
B. Robertson, Phys. Rev. Lett. {\bf 27},  1545  (1971).

\bibitem{MoTaNe2012}
P.~J. Mohr, B.~N. Taylor, and D.~B. Newell, Rev. Mod. Phys. {\bf 84},  1527
  (2012).

\bibitem{Ro1976}
G. Rosen, Phys. Rev. D {\bf 13},  830  (1976).

\bibitem{We1919}
H. Weyl, Ann. Phys. (Berlin) {\bf 364},  101  (1919).

\bibitem{Di1938LNH}
P.~A.~M. Dirac, Proc. Roy. Soc. London, Ser. A {\bf 165},  199  (1938).

\bibitem{Ed1931}
A.~S. Eddington, Proc. Camb. Phil. Soc. {\bf 27},  15  (1931).

\bibitem{Ed1924}
A.~S. Eddington, {\em \relax{The Mathematical Theory of Relativity}} (Cambridge
  University Press, Cambridge, England, 1924).

\bibitem{BuEtAl1997}
D.~L. Burke, R.~C. Field, G. Horton-Smith, J.~E. Spencer, D. Walz, S.~C.
  Berridge, W.~M. Bugg, K. Shmakov, A.~W. Weidemann, C. Bula, K.~T. McDonald,
  E.~J. Prebys, C. Bamber, S.~J. Boege, T. Koffas, T. Kotseroglou, A.~C.
  Melissinos, D.~D. Meyerhofer, D.~A. Reis, and W. Ragg, Phys. Rev. Lett. {\bf
  79},  1626  (1997).

\bibitem{BaEtAl1999}
C. Bamber, S.~J. Boege, T. Koffas, T. Kotseroglou, A.~C. Melissinos, D.~D.
  Meyerhofer, D.~A. Reis, W. Ragg, C. Bula, K.~T. McDonald, E.~J. Prebys, D.~L.
  Burke, R.~C. Field, G. Horton-Smith, J.~E. Spencer, D. Walz, S.~C. Berridge,
  W.~M. Bugg, K. Shmakov, and A.~W. Weidemann, Phys. Rev. D {\bf 60},  092004
  (1999).

\bibitem{Ca1953}
H.~B.~G. Casimir, Physica {\bf 19},  846  (1953).

\bibitem{Bo1968}
T.~H. Boyer, Phys. Rev. {\bf 174},  1764  (1968).

\bibitem{CaPo1948}
H.~B.~G. Casimir and D. Polder, Phys. Rev. {\bf 73},  360  (1948).

\bibitem{As1992}
A.~K.~T. Assis, Can. J. Phys. {\bf 70},  330  (1992).

\bibitem{As1995}
A.~K.~T. Assis,  in {\em Advanced Electromagnetism - Foundations, Theory and
  Applications}, edited by T.~W. Barrett and D. Grimes (World Scientific,
  Singapore, 1995), pp.\ 334--331.

\bibitem{Sa2000grg}
A.~D. Sakharov, Gen. Relativ. Gravit. {\bf 32},  365  (2000).

\bibitem{Vi2002}
M. Visser, Mod. Phys. Lett. A {\bf 17},  977  (2002).

\bibitem{On2013}
R. Onofrio, Mod. Phys. Lett. A {\bf 28},  1350022  (2013).

\bibitem{On2013epl}
R. Onofrio, Europhys. Lett. {\bf 104},  20002  (2013).

\bibitem{PoEtAl2010}
R. Pohl, A. Antognini, F. Nez, F.~D. Amaro, F. Biraben, J.~M.~R. Cardoso, D.~S.
  Covita, A. Dax, S. Dhawan, L.~M.~P. Fernandes, A. Giesen, T. Graf, T.~W.
  H\"{a}nsch, P. Indelicato, L. Julien, C.-Y. Kao, P. Knowles, E.-O. Le~Bigot,
  Y.~W. Liu, J.~A.~M. Lopes, L. Ludhova, C.~M.~B. Monteiro, F. Mulhauser, T.
  Nebel, P. Rabinowitz, J.~M.~F. dos Santos, L.~A. Schaller, K. Schuhmann, C.
  Schwob, D. Taqqu, J.~F. C.~A. Veloso, and F. Kottmann, Nature (London) {\bf
  466},  213  (2010).

\bibitem{AnEtAl2013}
A. Antognini, F. Nez, K. Schuhmann, F.~D. Amaro, F. Biraben, J.~M.~R. Cardoso,
  D.~S. Covita, A. Dax, S. Dhawan, M. Diepold, L.~M.~P. Fernandes, A. Giesen,
  A.~L. Gouvea, T. Graf, T.~W. H\"{a}nsch, P. Indelicato, L. Julien, C.-Y. Kao,
  P. Knowles, F. Kottmann, E.-O. Le~Bigot, Y.-W. Liu, J.~A.~M. Lopes, L.
  Ludhova, C.~M.~B. Monteiro, F. Mulhauser, T. Nebel, P. Rabinowitz, J.~M.~F.
  dos Santos, L.~A. Schaller, C. Schwob, D. Taqqu, J.~F. C.~A. Veloso, J.
  Vogelsang, and R. Pohl, Science {\bf 339},  417  (2013).

\bibitem{Je2011aop2}
U.~D. Jentschura, Ann. Phys. (N.Y.) {\bf 326},  516  (2011).

\bibitem{Ge1962}
M.~E. Gertsenshtein, Sov.~Phys.~JETP {\bf 14},  84  (1962).

\bibitem{ZeNo1983apsrmp}
Y.~B. Zeldovich and I.~D. Novikov, {\em The structure and evolution of the
  universe}, Rel.~Astrophys. Vol.~2, Chicago University Press, 1983.

\bibitem{AbHa1986}
L.~F. Abbott and D.~D. Harari, Nucl. Phys. B {\bf 264},  487  (1986).

\bibitem{BaSh2005}
F. Bastianelli and C. Schubert, J. High Energy Phys. {\bf 0502},  069  (2005).

\bibitem{BaNuScVi2007}
F. Bastianelli, U. Nucamendi, C. Schubert, and V.~M. Villanueva, J. High Energy
  Phys. {\bf 0711},  099  (2007).

\bibitem{BaSh2007}
F. Bastianelli and C. Schubert, J. High Energy Phys. {\bf 0903},  086  (2009).

\bibitem{BaCoDaSc2012}
F. Bastianelli, O. Corradini, J.~M. Davila, and C. Schubert, Phys. Lett. B {\bf
  716},  345  (2012).

\bibitem{RoWi2006}
S.~P. Robinson and F. Wilczek, Phys. Rev. Lett. {\bf 96},  231601  (2006).

\bibitem{BoIn1934}
M. Born and L. Infeld, Proc. Roy. Soc. London, Ser. A {\bf 144},  425  (1934).

\bibitem{In1936}
L. Infeld, Nature (London) {\bf 137},  658  (1936).

\bibitem{RaSt1988}
G. Raffelt and L. Stodolsky, Phys. Rev. D {\bf 37},  1237  (1988).

\bibitem{La1955apsrmp}
L.~D. Landau, {\em Niels Bohr and the Development of Physics}, edited by W.
  Pauli (McGraw--Hill, New York), p.~52., 1955.

\bibitem{TeEtAl1976}
H. Terazawa, K. Akama, and Y. Chikashige, Prog. Theor. Phys. (Kyoto) {\bf 56},
  1935  (1976).

\bibitem{TeEtAl1977}
H. Terazawa, Y. Chikashige, K. Akama, and T. Matsuki, Phys. Rev. D {\bf 15},
  1181  (1977).

\bibitem{KaLeTy1986}
H. Kawai, D.~C. Lewellen, and S.-H.~H. Tye, Nucl. Phys. B {\bf 269},  1
  (1986).

\bibitem{Gi1997}
G.~T. Gillies, Rep. Prog. Phys. {\bf 60},  151  (1997).

\bibitem{Mo2014}
P.~J. Mohr, private communication, 2014.

\bibitem{Br1992ieee}
J.~E. Brandenburg, IEEE Trans. Plasma Science {\bf 20},  944  (1992).

\bibitem{JeNo2013pra}
U.~D. Jentschura and J.~H. Noble, Phys. Rev. A {\bf 88},  022121  (2013).

\bibitem{Je2013}
U.~D. Jentschura, Phys. Rev. A {\bf 87},  032101  (2013), [Erratum Phys.~Rev.~A
  {\bf 87}, 069903(E) (2013)].

\bibitem{JeNo2014jpa}
U.~D. Jentschura and J.~H. Noble, J. Phys. A {\bf 47},  045402  (2014).

\bibitem{Je2014dirac}
U.~D. Jentschura, Ann. Phys. (Berlin) {\bf 526},  A47  (2014).

\bibitem{Je2014pra}
U.~D. Jentschura, Phys. Rev. A {\bf 90},  022112  (2014).

\end{thebibliography}
\end{document}